# Bias and sensitivity analysis for unmeasured confounders in linear structural equation models

Adam J. Sullivan and Tyler J. VanderWeele

Abstract. In this paper, we consider the extent of the biases that may arise when an unmeasured confounder is omitted from a structural equation model (SEM) and we propose sensitivity analysis techniques to correct for such biases. We give an analysis of which effects in an SEM are, and are not, biased by an unmeasured confounder. It is shown that a single unmeasured confounder will bias not just one, but numerous, effects in an SEM. We present sensitivity analysis techniques to correct for biases in total, direct, and indirect effects when using SEM analyses, and illustrate these techniques with a study of aging and cognitive function.

# 1 Introduction

Linear structural equation models(LSEMs) are frequently used in many of the social sciences Bollen (1989), Sobel (1990). Many of the models used are complex and contain various interrelationships. These relationships are shown in a model as paths. LSEMs allow researchers to analyze multiple paths at the same time. However it is often easy to think of variables that have been left out of a model which may have an impact on these relationships. This impact comes in the form of biasing the effects which prompts the need for sensitivity analysis. There is a large literature on sensitivity analysis for unmeasured confounding for a single cause-effect relationship Vanderweele (2008), Vanderweele and Arah (2011), Kitagawa (1955), Cornfield, Haenszel, Hammond, Lilienfeld, Shimkin, Wynder, Cornfield, Haenszel, Hammond, Lilienfeld, Shimkin, and Wynder (2009), Breslow and Day (1980), Schlesselman (1978), Bross (1966), Rosenbaum and Rubin (1983), Yanagawa (1984), Gail, Wacholder, and Lubin (1988), Flanders and Khoury (1990), Copas and Li (1997), Lin, Psaty, and Kronrnal (1998a), Robins, Rotnitzky, and Scharfstein (2000), Rosenbaum (2002), Imbens (2003), Greenland (2003, 2005), Brumback, Hernán, Haneuse, and Robins (2004), Stürmer, Schneeweiss, Avorn, and Glynn (2005), McCandless, Gustafson, and Levy (2007), Arah, Chiba, and Greenland (2008), Lash, Fox, and Fink (2011), Lin, Psaty, and Kronmal (1998b), Imai, Keele, and Yamamoto (2010). Here we apply and extend this literature to the setting of LSEMs with many cause-effect relationships and a single unmeasured confounder.

A confounder is an extra variable that is causally related to both the dependent and the independent variable. A confounder may be called an unmeasured confounder if either no data was collected on it or it was left out of the model that was analyzed. LSEMs have very strong assumptions that are made about the functional form of the relationships, the distribution of variables and having included all confounding variables in the model. Many of the assumptions have been ignored in the models when used in practice VanderWeele (2012). With the strong assumptions about confounding it is important to know how sensitive the effects of interest are with respect to unmeasured confounding.

In this paper we will describe what a LSEM is and discuss the basic assumptions. Then we will consider sensitivity analysis where we will show in what circumstances and for which effects an unmeasured confounder would bias the results and what estimates are robust to the bias. We will then discuss what other effects aside from the effects of interest are biased due to unmeasured confounding. We will then give an example and discuss how to use this sensitivity analysis technique. Then we will consider a missing path analysis in which we will discuss how the bias is affected if certain paths are absent from the model.

## 2 Brief overview of Linear Structural Equation Models

We will begin by considering a brief overview of the basics of an LSEM. We give an overview of the important concepts and language as well as a background to the methods contained in the rest of this paper.

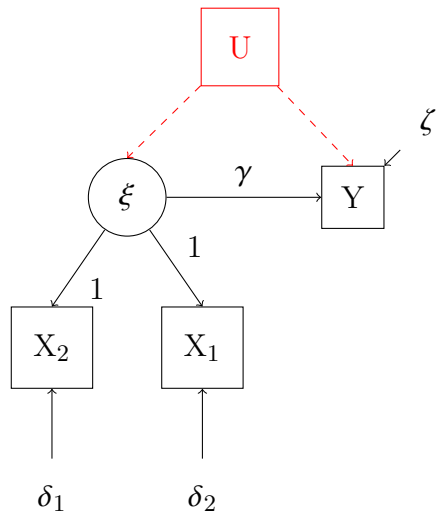

Figure 1: Path Diagram Example from Bollen with Newly Added U

Figure 1 is a simple path diagram that is shown by Bollen Bollen (1989). The model contains both observable and latent variables. Observable variables are variables that have been directly measured and are represented by squares in the figure. Latent variables are variables that are not able to be directly measured but can be inferred from observable variables in the model and are represented by circles in the figure. Relationships between these variables are represented by the arrows which are called paths. A line with a single arrow represents a causal path, for example in figure 1, the arrow $\xi \to Y$, represents that $\xi$ has a direct effect on Y.

The final parts of figure 1 which have not been mentioned previously are $\delta_1$, $\delta_2$, and $\zeta$. These represent random error effects on $X_1$, $X_2$ and Y respectively. Random error is the element of randomness that is not contributed by the other paths in the model. It should also be noted that unless otherwise indicated, these errors are all uncorrelated with each other and with $\xi$. In a LSEM, correlations are shown by using a double headed arrow path between two variables. In the rest of the path diagrams in this paper we will assume the errors are uncorrelated and leave the error terms out for simplicity. However we note that there is still error associated with each of the relationships as in figure 1. From the complete model in Figure 1 a LSEM assumes the following mathematical relationships:

$$\begin{aligned} y &= \gamma \xi + \zeta \\ x_1 &= \xi + \delta_1 \\ x_2 &= \xi + \delta_2 \end{aligned}$$

Traditionally LSEM assume that all of the above equations involve variables that are normally distributed as well as that each path follows a linear regression. Various extensions are possible for binary and ordinal variables in which it is assumed that the observed binary or categorical variables is the dichotomized or coarsened version of an underlying latent continuous normally distributed variable Bollen (1989).

Suppose now we have a variable U in the model that represents a missing variable that was not accounted for in the analysis. This would bias our effect estimates for the effect of $\xi$ on Y. This is also sometimes called an unblocked "backdoor path" Pearl (2000). If that variable had not been accounted for then even if there was no effect of $\xi$ on Y (i.e. the arrow between the variables is missing) we would likely find an association between $\xi$ and Y because of U. We would likely believe there is a causal relationship between $\xi$ and Y even though there is no true association present. We need to be very mindful of this when positing models and this paper will detail steps to take in order to assess how sensitive results on an LSEM are to the impact of an unmeasured U. We will consider sensitivity analysis for total effects and for direct and indirect effects which arise when multiple variables and cause-effect relationships are being considered.

We note that in some settings an unmeasured confounder can be explicitly included in a LSEM and in certain settings it is still possible to proceed with estimation of certain effects Bollen (1989). Here we deal with the setting when the unmeasured confounder has in fact not been included in the model and the researcher is interested in how the unmeasured confounder biases effect estimates.

# 3 Confounding and Bias of Effects

## 3.1 Path Analysis with Confounding

Whether or not there is bias for a specific effect in the context of a LSEM, will depend on whether all of the "backdoor paths" between the exposure and outcome are blocked or controlled for. In this section we will determine whether there is bias within the effect estimates that the LSEM has given us due to an incorrect model with a missing confounder.

We will consider what happens when a model is missing a single confounder. We will first consider for which paths a potential confounder might cause bias for the effect estimates. To illustrate this we will use figure 2. Suppose, as in figure 2 we have A which is an exposure, M which is a mediator, and Y which is an outcome of interest. On a more complex diagram, A, M and Y could be any three variables on the diagram for which we were interested in the various total, direct and indirect effects. Once we have chosen these three variables the analysis of bias can proceed as described below. If we change the three variables chosen as the exposure, mediator and outcome, we could proceed with a similar analysis for these three variables as well. C1 and C2 are measured covariates and U is again an unmeasured confounder. Note that in our models all of the variables are observable; however any of these variables could be latent as well with multiple observed indicators as is typical in LSEM and this would not change the results nor the sensitivity analysis. We will illustrate the results with latent variables with multiple observed indicators in Section 6 below.

The direct effect is the path from the exposure, A, to the outcome, Y, which is not mediated by any other variable. This is represented by the $A \to Y$ path in figure 2. An indirect effect is a path from A to Y that goes through one or more other variables. This is referred to as the effect being mediated by other variables. This is represented by the path, $A \to M \to Y$ in figure 2. The total effect is the combination of the direct and the indirect effect. The total effect of $A$ on $Y$ would include both the direct path from $A \to Y$ as well as the path $A \to M \to Y$.

## 3.2 Bias of Causal Effects

In this section we will consider diagrams similar to figure 2. We will change the location of $U$ in the diagram and examine how this affects the bias of the three types of effects discussed in section 3.1 (total, direct and indirect).

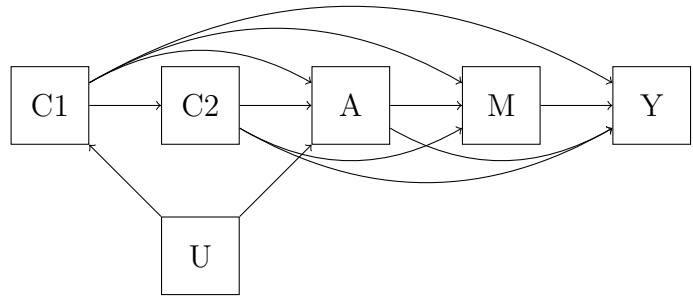

Figure 2: Example SEM

We summarize the results in Table 1 and formal arguments are given in the appendix. Table 1 is laid out so that the rows represent where the left arrow of the U variable points on the pathway from the treatment to the outcome and the columns represent where the right arrow from U points on the pathway.

Table 1: Bias of Causal Effects

| | | Total Effects | | | | |
|---|---|---|---|---|---|---|
| | | Right | | | | |
| | | Pre-Exposure | Exposure | Mediator | Outcome | Post-Outcome |
| Left | Pre-Exposure | Not Biased | Not Biased | Not Biased | Not Biased | Not Biased |
| | Exposure | | | Biased | Biased | Not Biased |
| | Mediator | | | | Not Biased | Not Biased |
| | Outcome | | | | | Not Biased |
| | Post-Outcome | | | | | Not Biased |

| | | Direct / Indirect Effects | | | | |
|---|---|---|---|---|---|---|
| | | Right | | | | |
| | | Pre-Exposure | Exposure | Mediator | Outcome | Post-Outcome |
| Left | Pre-Exposure | Not Biased | Not Biased | Not Biased | Not Biased | Not Biased |
| | Exposure | | | Indirect Biased | Direct Biased | Not Biased |
| | Mediator | | | | Direct and Indirect Biased | Not Biased |
| | Outcome | | | | | Not Biased |
| | Post-Outcome | | | | | Not Biased |

The cells indicate which of the effects are biased and which are unbiased in each of the various settings. Justification is given in the appendix. Cells in the table with the left arrow indicated to the right of the right arrow were left blank. In most scenarios considered in Table 1, the unmeasured variable would not produce bias for the total, indirect and direct effect of A on Y. This is because many of the scenarios in Table 1 the unmeasured variable affects two variables which occur either before the exposure or after the outcome and thus does not bias the effects of interest. Table 1 gives only 3 scenarios which results in a bias of the effects of interest. Figure 3 shows the three interesting scenarios. The first one shown in figure 3a is a case of exposure-mediator confounding. This is confounding where the U has a causal relationship with both the exposure and with the mediator. With this scenario there is bias in the total effect as well as the indirect effect. There is a possibility of an association between A and M purely due to the unmeasured U. This would mean that there may in fact be no indirect effect of $A \to M \to Y$ but our estimates would show otherwise. The direct effect however is unbiased in this scenario. The second interesting scenario is shown in figure 3b, this is a case of exposure-outcome confounding, meaning that the unmeasured confounder has a causal relationship with the exposure and the outcome. With this scenario there is bias in the total and the direct effect; but the indirect effect is unbiased. Finally the last interesting scenario is shown in figure 3c. This is a case of mediator-outcome confounding. In this scenario both the direct and indirect effects are biased. In the next section we will explore the bias that is present in these scenarios and develop sensitivity analysis that can be used to assess the impact of unmeasured variables.

The variables in the models we have considered are very simple. However one can apply the results shown here to a LSEM of any size and complexity. By breaking down the complex model into smaller parts like these scenarios one can assess for bias due to confounding in different parts of the model. If there are intermediate variables between the variables in the diagram chosen as the exposure A and mediator M and one of the arrows of U is pointed into one of these intermediate variables, the bias analysis would be analogous to the setting in which the arrow of U pointing into M itself. If there are intermediate variables between the variables in the diagram chosen as the mediator M and outcome Y and one of the arrows of U pointed into one of the intermediate variables, the bias analysis would be analogous to the arrow of U pointing into Y.

Each of the variables in figure 3 are displayed as a single variable for the purpose of simplicity. However each of these variables can represent a group of variables. For example in each of scenarios the variable C can be used to represent all of the covariates that are being adjusted for. An example of this variable grouping will be shown in section 6.

# 4 Scope of Bias throughout a Structural Equation Model

One key feature of an LSEM is the capacity to estimate the effects for any path specified in the model. In section 3 we considered the bias created by unmeasured confounding on the main effects of interest. This section will consider what other effects in the LSEM are biased due to the unmeasured confounding. We will show that with unmeasured confounding not just one edge and effect estimate are biased but in fact many effect estimates and even numerous distinct edges will be biased. Specifically, we show in the appendix that for any variable V that has an edge into the variable at the left most edge of the unmeasured confounder, the effect estimate of the edge from V to the variable on the right most edge of the unmeasured confounder will be biased. This means in scenario 1 any effect estimate of edges into M will be biased for any variable that affects A. In scenarios 2 and 3 any effect estimate of edges into the outcome, Y, will be biased for any variable that affects A in scenario 2 or that affects M in scenario 3.

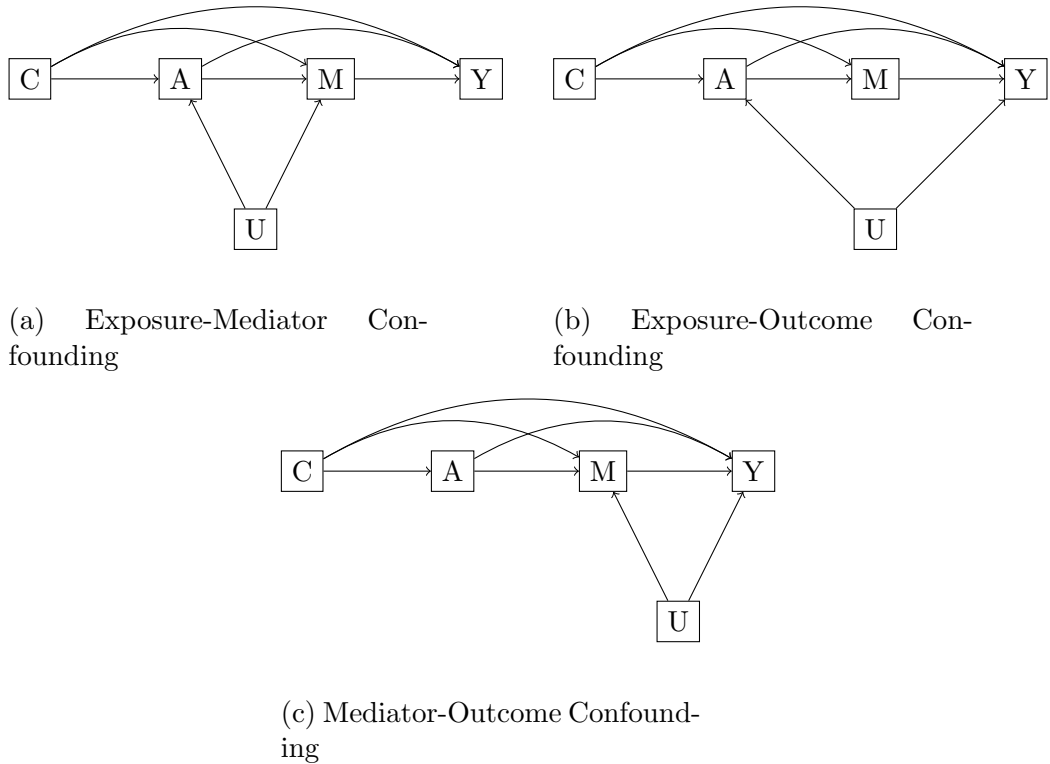

(a) Exposure-Mediator Confounding

(b) Exposure-Outcome Confounding

(c) Mediator-Outcome Confounding

Figure 3: Scenarios with Bias Present

In order to further explore bias we will consider scenario 3, mediator-outcome confounding, in more detail using simulations. The other two scenarios will be considered in the appendix. Recall from section 3.2 that the M → Y relationship as well as the A → Y relationship is known to be biased.

For example, simulations can be used to illustrate that the C → Y relationship is also biased. The following steps were done to simulate the data which follows the paths shown in figure 3c:

1. C and U are normally distributed with mean 0 and standard deviation 1.
2. 20,000 values are simulated for both C and U.
3. The effects of C→ A, C→ M, A→ M, A→ Y, U→ M, U→ Y and M→ Y were all set to a moderate effect of 0.6.
4. The effect of C→ Y was set to 0.
5. A = 0.6C + $\epsilon_a$, where $\epsilon_a$ is a random error term normally distributed with mean 0 and standard deviation 1.
6. M = 0.6C + 0.6A + 0.6U + $\epsilon_m$, where $\epsilon_m$ is a random error term normally distributed with mean 0 and standard deviation 1.
7. Y = 0C + 0.6A + 0.6M + 0.6U + $\epsilon_y$, where $\epsilon_y$ is a random error term normally distributed with mean 0 and standard deviation 1.

This defines the paths exactly as in figure 3c and gives us a data set with 5 variables and 20,000 values for each variable. Then using Stata 13 the SEM was fit two ways. The first fit was having the C → Y edge in the model, knowing that this has a true effect of 0. The second model fit was leaving the C → Y edge out of the model. Results are summarized in Table 2.

| | Scenario 3: Mediator-Outcome Confounding | | | | | |
|---|---|---|---|---|---|---|
| | C → A | C → M | A → M | C → Y | A → Y | M → Y |
| True Model | 0.6 | 0.6 | 0.6 | 0 | 0.6 | 0.6 |
| C → Y in model | 0.59 (0.58, 0.61) | 0.61 (0.59, 0.63) | 0.58 (0.56, 0.60) | -0.17 (-0.19, -0.15) | 0.44 (0.42, 0.45) | 0.87 (0.86, 0.88) |
| C → Y not in model | 0.60 (0.58, 0.61) | 0.61 (0.59, 0.63) | 0.58 (0.56, 0.60) | – | 0.40 (0.39, 0.42) | 0.82 (0.81, 0.83) |

Table 2: SEM effect estimates from scenario 3 simulation.

When the model is fit allowing for the possibility of a *C → Y* edge (first row) the *M → Y* effect is upwards biased, with an estimate of 0.87 with or% confidence interval (0.86, 0.88) instead of 0.6; the *A → Y* effect is biased downwards with an estimate of 0.44 (0.42, 0.45) instead of 0.6. We also find that the effect of *C → Y* is biased downward with an estimate of -0.17 (-0.19, -0.15) instead of 0. Again this effect was set to zero in the simulations. When leaving the *C → Y* path out of the model the effect of *A → Y* and *M → Y* remain biased downwards and biased upwards respectively. The bias of the *A → Y* edge is even larger than before with estimates of 0.40 (0.39, 0.42).

Our simulations in Table 2 illustrates that more effects other than the direct and indirect effect are biased. An unmeasured confounder does not just bias a single edge but many edges. A single unmeasured confounder can thus introduce bias for many paths in an LSEM. Correcting biases for a LSEM single unmeasured confounder does not just require correcting one edge but many.

We will give methods in the next section for sensitivity analysis for specific total, direct or indirect effects of interest. But when using such sensitivity analysis it is recommended that in the presence of unmeasured confounding researchers also mention the possibility of the other effects that could be biased significantly as well.

# 5 Sensitivity Analysis for Structural Equation Models Under Unmeasured Confounding

## 5.1 Scenario 1: Exposure - Mediator Confounding

Consider the scenario in Figure 3a from Table 1. Here we see that the total and indirect effects both are biased. The bias comes from the possibility of an association being present based on the unblocked "backdoor" path that was mentioned in section 2. Our goal is to quantify the amount of bias that is present and to assess the robustness of the effects to confounding with sensitivity analysis. Formal mathematical development is given in the appendix (Section 9).

### 5.1.1 Total Effect Bias and Correction.

Consider the total effect under exposure-mediator confounding. This confounding will bias the $A \to M$ relationship and this in turn biases the total effect for the $A \to Y$ relationship. We will consider sensitivity analysis in this setting. Let $B_{add}$ denote the difference between the quantity estimated by the LSEM (ignoring $U$) and the true causal effect of $A$ on $Y$ (i.e. what we would have obtained had we included $U$ as well). Suppose that the effect of $U$ on $Y$ is constant across strata of $A$ (i.e. $U$ and $A$ do not interact in their effects on $Y$, this is typically assumed in a LSEM) and the mean of $U$ is additive in both $C$ and $A$, VanderWeele VanderWeele (2010) and Lin Lin et al. (1998b) have shown that the bias is given by

$$B_{add} = \gamma d$$

where $\gamma$ is the mean effect of $U$ on $Y$, and $d = m_1 - m_0$, where $m_0$ and $m_1$ are the means of $U$ for two different levels of $A$ being compared. If $U$ were binary $m_0$ and $m_1$ would be the prevalences of $U$ in the two different levels of $A$ being compared. The parameter $\gamma$ is the estimated effect that we would see if we regressed $Y$ on $U$ and $M$. Once both $\gamma$ and $d$ are specified we can then subtract the bias $\gamma d$ from the estimate of the effect of $A$ on $Y$ from the LSEM to get a corrected estimate of the effect of $A$ on $M$. We can also subtract the bias factor $\gamma d$ from both limits of the 95% confidence intervals in order to also get a corrected 95% confidence interval.

In general it is helpful to vary the values specified for both $\gamma$ and $d$, as this will allow one to assess the sensitivity of the estimate of the effect of $A$ on $Y$. In most cases the true values of $\gamma$ and $d$ are unknown and need to be specified. We vary our specifications for both $\gamma$ and $d$. This leads us to having a range of values for the bias, $B_{add}$. Subtracting this range of values from the estimate of the total effect we then have a range of values for the correct total effect. We can then take the range of values for the bias, $B_{add}$, and subtract them from the upper and lower bounds of the confidence interval around the total effect in order to correct the confidence intervals as well.[1]

### 5.1.2 Indirect Effect Bias and Correction.

In section 3, we noted that the direct effect is unbiased by exposure-mediator confounding so we do not need to do sensitivity analysis for this effect. However, the indirect effect is biased by exposure-mediator confounding. In section 2 it was noted that the total effect is the combination of the direct and indirect effects. Since the direct effect is unbiased and since

$$\text{Total Effect} = \text{Direct Effect} + \text{Indirect Effect}$$

the bias for the indirect effect will be the same as the bias for the total effect. Once the sensitivity analysis parameters $\gamma$ and $d$ are specified, we can thus take the same bias factor $\gamma d$ and subtract this from the indirect effect and both limits of its confidence interval to get a corrected estimate and confidence interval for the indirect effect. We can use this approach to assess the sensitivity of the indirect effect. As we vary the parameters, we can take the range of values for the bias factor of the total effect used previously and subtract them from the indirect effect and its confidence interval. This again gives us a corrected indirect effect as well as corrected confidence interval in each case.

---

[1] Alternatively, it would have been possible to use sensitivity analysis to correct the effect of $A$ on $M$ first and then use this to get a corrected total effect as well, but obtaining corrected confidence intervals by correcting the $A \rightarrow M$ relationship first is considerably more complicated, whereas the correction approach presented here for estimates and confidence intervals is relatively straightforward and as will be seen below we will likewise be able to obtain similar corrected estimates and confidence intervals for the direct and indirect effects as well.

## 5.2 Scenario 2: Exposure - Outcome Confounding

Consider the scenario in Figure 3b from Table 1 of unmeasured exposure-outcome confounding. Here we see that both the direct and total effects are biased. As noted in Table 1 the indirect effect is unbiased.

### 5.2.1 Direct effect Bias and Correction.

Examining the direct effect we see that the unmeasured confounding biases the direct effect $A \to Y$ relationship. For sensitivity analysis for the direct effect we can proceed in a similar manner as in Section 5.2. If $U$ is additive in both $C$ and $A$ and the effect of $U$ on $Y$ is constant across strata of $A$ then if we define $B_{add}$ as the difference between our estimator using the LSEM ignoring the unmeasured confounder and the true direct effect, then we have that:

$$B_{add} = \gamma d,$$

where $\gamma$ is the mean effect of $U$ on $Y$ and $d = (m_1 - m_0)$, where $m_1$ and $m_0$ are the mean of $U$ for the two different levels of the exposure $A$ being compared. We again specify values for $\gamma$ and $d$ and calculate $B_{add} = \gamma d$. This represents the bias for the direct effect of $A$ on $Y$. We can again subtract this bias factor from our estimate of the direct effect and both limits of its confidence interval to obtain a corrected estimate and confidence interval. We can then vary the values of both $\gamma$ and $d$ and obtain a range of values and confidence intervals for the direct effect to assess its sensitivity to unmeasured confounding.

### 5.2.2 Total Effect Bias and Correction.

As discussed in section 3, the indirect effect is not biased by unmeasured exposure-outcome confounding, but the total effect is biased by such confounding. As in section 5.1.1 we know that

$$\text{Total Effect} = \text{Direct Effect} + \text{Indirect Effect}.$$

Since the indirect effect is unbiased, for any given level of the sensitivity parameters $\gamma$ and $d$, we can take the bias factor for the direct effect found previously and subtract this from the total effect and both limits of its confidence interval to obtain a corrected total effect and confidence interval. We can do this for a range of values of the sensitivity analysis parameters for the total effect as well.

## 5.3 Scenario 3: Mediator - Outcome Confounding.

Consider the scenario in Figure 3c from Table 1 with unmeasured mediator-outcome confounding. Here we see that the direct and indirect effects are both biased by such unmeasured mediator-outcome confounding. The total effect in this scenario is however unbiased.

### 5.3.1 Direct Effect Bias and Correction.

Consider the direct effect of $A$ on $Y$. If we are able to assume the direct effect of $U$ on the outcome $Y$ is the same for all levels of $a$ (i.e. no interaction between $U$ and $A$) and the expected value of $U$ is additive in $A$ and $(M,C)$ (i.e. $E[U|a,m,c] = g(a) + h(m,c)$ for some functions $g$ and $h$) then if we define the bias factor $B_{add}$ to be the difference between our estimator of the direct effect and the true direct effect we have Vanderweele and Arah (2011):

$$B_{add} = \delta\gamma$$

where $\delta$ is the difference in the mean value of $U$, conditional on $M$, between the two levels of $A$ we are comparing and $\gamma$ is the direct effect of $U$ on the outcome $Y$, not through $M$

As before we will need to specify both $\delta$ and $\gamma$. Once we have specified a range of values for both $\delta$ and $\gamma$ we have a range of values for $B_{add}$. To assess the sensitivity of the direct effect we subtract $B_{add}$ from the estimate and both limits of the confidence interval. We then have obtained a corrected estimate and confidence interval for the direct effect.

### 5.3.2 Indirect Effect Bias and Correction.

We noted in section 3 that the total effect is unbiased by unmeasured mediator-outcome confounding. Since

$$\text{Total Effect} = \text{Direct Effect} + \text{Indirect Effect}.$$

the bias for the indirect effect will be of equal magnitude but opposite sign as that of the direct effect. We can thus add the bias factor for the direct effect to the estimate and both limits of the confidence interval of the indirect effect to get a corrected indirect effect. We can do this for a range of values for the sensitivity analysis parameters so that we can assess the sensitivity of the indirect effect to the unmeasured confounding.

## 5.4 Discussion of Sensitivity Findings.

We now will consider the range of values for our effects once we have specified a range of values for the sensitivity parameters. Three things may happen with these ranges:

1. The range of the effects will contain zero.
2. The range of the effects will be in the same direction as the results of the data analysis.
3. the range of the effects is in the opposite direction as the results of the data analysis.

If the range of the effect contains zero this means that it is possible that the effect of interest is sensitive to unmeasured confounding. We would have evidence that the effect which was seen before sensitivity analysis may be due to the unmeasured confounding.

If the range of values is in the same direction of the effect originally found and does not include zero, this would indicate that the effect is relatively robust to unmeasured confounding at least over the range of sensitivity parameters considered.

If part of the range of the values is in the opposite direction of the effect originally found then it is possible that the confounding is strong enough to have changed the direction of the effect.

In some cases, for an estimate effect to be reduced to zero, very large values of the sensitivity analysis parameters may be required. This then would provide evidence that in fact the effect under study is actually present, and not entirely due to unmeasured confounding alone.

# 6 An Example

In this section we give an example of the use and interpretation of a sensitivity analysis in the context of a published study on cognitive function. This example comes for the work of Charlton et al Charlton, Landau, Schiavone, Barrick, Clark, Markus, and Morris (2008). This paper considered the relationship between age and working memory. Charlton et al Charlton et al. (2008) also evaluated whether DTI(diffusion tensor imaging) measured white matter mediated the relationship between age and information processing speed, working memory, flexibility and fluid intelligence. Charlton et al Charlton et al. (2008) contained 118 subjects ages 50-90 with mean age 70 and standard deviation 10.5. Processing speed, working memory, flexibility and fluid intelligence were assessed by standardized neuropsychological tests (see table 1 in Charlton et al for more information on these tests). Based on their study Penke & Deary Penke and Deary (2010) proposed the model which is shown in figure 4. This model contains one outcome called the general factor of cognitivity or g factor which is a latent variable and is inferred by processing speed, working memory, flexibility and fluid intelligence. Their model includes a direct effect of age on the g-factor outcome as well as an effect mediated by DTI mean diffusivity.

Figure 4 here represents figure 1 in Penke & Deary with an added unmeasured confounder of the relationship DTI mean diffusivity and the g factor. There is the possibility, for example, of a genetic or biological factor that would lead to an increase in DTI mean diffusivity(decrease in white matter integrity Charlton et al. (2008)) and also a decrease in g factor. For the model fit by Penke and Deary, as indicated in Figure 4, the direct effect of age on g factor was -0.65 (-0.67, -0.62); the indirect effect of age on g factor through DTI mean diffusivity is 0.0077 (0.0077, 0.0078). The indirect effect is found by multiplying the effect of age on DTI mean diffusivity by the effect of DTI mean diffusivity on g factor ($0.77 * 0.01 = 0.0077$). All of the coefficients have been standardized, meaning that the original variables in the model were transformed into variables with mean 0 and standard deviation 1. We will use the sensitivity analysis outlined in section 5.3 to consider how an unmeasured mediator-outcome confounder might change these estimates

We begin by assessing the sensitivity of the direct effect. This has a bias of the form $B_{add}(m) = \delta\gamma$, where for a fixed level of DTI mean diffusivity= $m$, $\delta$ is the difference in prevalence of U between two ages one standard deviation apart and $\gamma$ is the effect of $U$ on the g factor. We begin with considering what values for $\delta$ and $\gamma$ would suffice to eliminate the direct effect. This could be done if the difference in prevalence of $U$ at fixed $m$ in ages one standard deviation apart was $\delta=0.13$, and the corresponding $\gamma$ was -5. This would lead to $B_{add} = (0.13)(-5) = -0.65$. If we subtracted this bias from the model's estimate of the direct effect, we would obtain $-0.65 - -0.65 = 0$. With a corrected 95% confidence interval of (-0.02,0.03), and this would suffice to completely eliminate the effect. Alternatively, a difference in prevalences of 0.26 with an effect of $U$ on $Y$ of 2.5 would likewise give a bias factor of $(0.26)(-2.5)=-0.65$ and suffice to eliminate the direct effect. While a difference in prevalences of 0.13 (or possibly even 0.26) might be considered plausible, an effect size of $U$ on $Y$ of 5 or even 2.5 standard deviations is probably unlikely. That such extreme values for $\gamma$ would be required to eliminate the effect suggests that the direct effect is reasonably robust to unmeasured confounding.

In contrast, with the indirect effect, a prevalence difference of $\delta = 0.05$ and an effect of $U$ on $Y$ of $\gamma = -0.154$ standard deviations would give a bias factor of $B_{add} = (0.05) - (0.154) = -0.0077$ which would suffice to explain away the indirect effect. This is a much more modest scenario than we considered for the direct effect. Note that, as in section 5.3, we add this bias factor to the indirect effect to get the corrected indirect effect, while we subtracted it from the direct effect to get the corrected direct effect. If instead we specified the prevalence difference to be $\delta=0.05$ and the effect of $U$ on $Y$ were $\gamma = 0.3$ standard deviations, this would would give a bias factor of $B_{add} = (0.05)(0.3) = 0.015$ and a corrected estimate and confidence interval of -.0073 (-.0073, -.0072) which was the opposite direction of the initial effect. We see that much less unnmeasured confounding is needed to eliminate the indirect effect than was the case with the direct effect. With the indirect effect even a fairly modest amount of confounding could reverse the direction of the effect. In this example we can be fairly confident that the direct effect is robust to unmeasured confounding, but we see also that, due to the possibility of unmeasured confounding we cannot really draw conclusions about the indirect effect.

To illustrate sensitivity analysis further, we could also try specifying $\gamma$ to be similar in magnitude to the effect of other variables on the g factor. For example, if we specified the effect of $U$ on $Y$ to be of the same magnitude as the effect of age on the g-factor, -0.65, and specified the prevalence difference to be $\delta=0.13$ once again, we would have $B_{add} = (0.13)(-0.65) = -0.0845$. When we subtract the bias factor from our estimated direct effect and both limits of the confidence interval, we get a corrected direct effect estimate of $-0.5655(-0.5855, -0.5355)$, which is not very different from the initial estimate.

When we add this bias factor to the indirect effect and both limits of its confidence interval we obtain $-0.0768(-0.0768, -0.0767)$, which again is the reverse direction of the initial indirect effect estimate.

We can also consider how unmeasured confounding would affect our direct and indirect effect estimates if the effect of $U$ on the g factor were exactly one standard deviation. If we keep $d = 0.13$ and we let $\gamma = -1$ we have a bias factor of $B_{add} = (0.13)(-1) = -0.13$, and the corrected direct effect would be $-0.52(-0.54, -0.49)$ and the corrected indirect effect would be $-0.1223(-0.1223, -0.1222)$. If we keep $d = 0.13$ and we let $\gamma = 1$ we have a bias factor of $B_{add} = (0.13)(1) = 0.13$, the corrected direct effect would be $-0.78(-0.8, -0.75)$ and the corrected indirect effect would be $0.1377(0.1377, 0.1378)$. With $\gamma = 1$ or $\gamma = -1$ these would be a fairly large genetic effects; however the direct effect still would not be eliminated, but the direction of the indirect effect is again reversed. Again the indirect effect is sensitive to unmeasured confounding. It is thus possible that there is no effect of DTI mean diffusivity on g factor and that any effect seen in the model is due to the unmeasured confounding.

In summary, our sensitivity analysis suggests that it is unlikely any unmeasured confounder of DTI mean diffusivity and g factor would explain away the direct effect of age on g factor. It also suggests, however, that the indirect effect of age on g factor through DTI mean diffusivity is highly sensitive to unmeasured confounding.

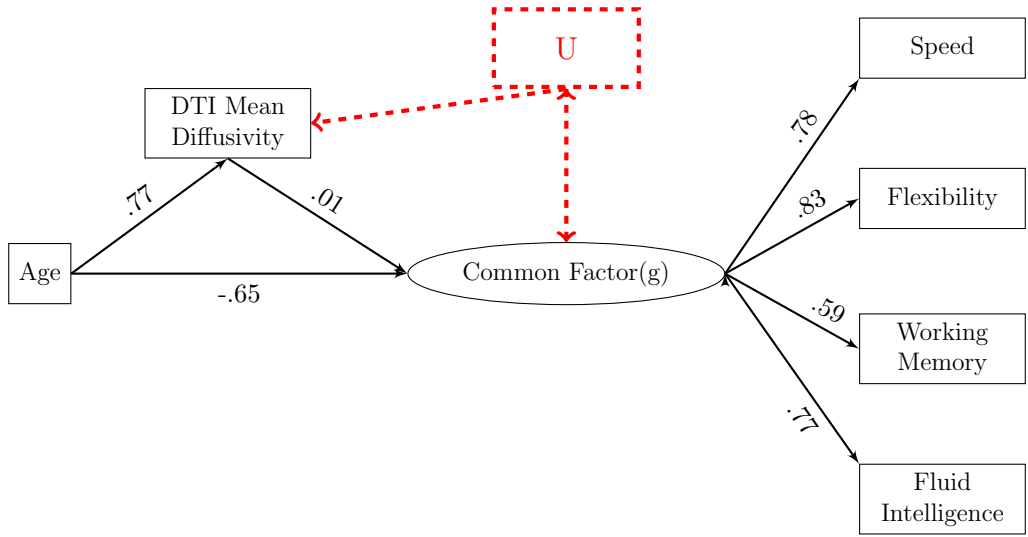

Figure 4: Penke and Deary: Figure 1 with Added Confounding

# 7 Missing Path Analysis

In this section we will consider what happens to the biases considered above when specific paths are absent. We will consider the three scenarios discussed in section 5 and remove one path at a time checking whether effects remain biased. Table 3 displays the results if specific edges are absent and removed from the model. The table is split into 3 sections for the total, direct and indirect effect. If we consider the total effect section of the table and then if the $A \to Y$ path is absent, we see that in the presence of exposure-mediator confounding the total effect is still biased. However had we been able to remove the $M \to Y$ path, the total effect would be unbiased. This means that if it is known that the $M \to Y$ path does not exist in the model (i.e. it is reasonable to assume that it has no effect) then the result would be that the total and direct effects are both unbiased. When removing either of these effects the indirect effect would be equal to zero.

This table will help the researcher to determine whether or not a sensitivity analysis is appropriate for a particular model. We note that removing a path should only be done if it is truly known on substantive grounds that it is absent. A key assumption that we have explored in this paper is that the LSEM is complete. This means we cannot remove a path that should be there without introducing further bias.

Table 3 corresponds to the biases if one of the arrows ($A \to M$, $A \to M$ or $M \to Y$) is missing but we have an unmeasured confounder not included in the model and we simply fit the SEM. However if we know for certain that one of the edges ($A \to M$, $A \to M$ or $M \to Y$) were missing then without fitting the SEM we could still estimate certain effects. For example, if there were an unmeasured exposure-mediator confounder and if we knew that the $A \to M$ edge were missing we would know the the indirect effect is 0 even though the SEM would not estimate as 0 if we ignored the unmeasured confounding.

| Total effect | | Missing Arrows | | |
|---|---|---|---|---|
| | | $A \to Y$ | $A \to M$ | $M \to Y$ |
| | Exposure-Mediator Confounding | Biased | Biased | Unbiased |
| | Exposure-Outcome Confounding | Biased | Biased | Biased |
| | Mediator-Outcome Confounding | Unbiased | Unbiased | Unbiased |

| Direct effect | | Missing Arrows | | |
|---|---|---|---|---|
| | | $A \to Y$ | $A \to M$ | $M \to Y$ |
| | Exposure-Mediator Confounding | Unbiased | Unbiased | Unbiased |
| | Exposure-Outcome Confounding | Biased | Biased | Biased |
| | Mediator-Outcome Confounding | Biased | Biased | Unbiased |

| Indirect effect | | Missing Arrows | | |
|---|---|---|---|---|
| | | $A \to Y$ | $A \to M$ | $M \to Y$ |
| | Exposure-Mediator Confounding | Biased | Biased | Unbiased |
| | Exposure-Outcome Confounding | Unbiased | Unbiased | Unbiased |
| | Mediator-Outcome Confounding | Biased | Unbiased | Biased |

Table 3: Missing Path Analysis

# 8 Discussion

This paper considers sensitivity analysis for LSEMs. Three scenarios, exposure-outcome confounding, exposure-mediator confounding and mediator-outcome confounding were found to have bias associated with either the total, direct or indirect effects, as well as potentially numerous others. Section 5 showed a straight forward sensitivity analysis for each of the three scenarios. We also showed the result that when you have an unmeasured confounder other effects in the model will be biased as well. Specifically, for any variable V that has an edge into the variable at the left most edge of the unmeasured confounder, the effect estimate of the edge from V to the variable on the right most edge of the unmeasured confounder will be biased. This is the case for all potential variables V and thus many edges may be biased by a single unmeasured confounder. For example in exposure-outcome confounding, any other edge into the outcome will have a biased effect estimate if that variable also affects the exposure. This paper then showed situations in which the bias of the primary effects of interest would be absent if it is known that certain edges in the model were absent. However this should only be done if there is a valid reason for assuming that a specific edge does not exist. Otherwise removing an edge will leave the model incorrectly specified and the results would be biased. Theoretical explanations for the work shown in this paper are discussed in the appendix.

We want to conclude by considering the importance and impact of what this kind of analysis can do. When unmeasured confounding is present we cannot be sure if our results are reliable. This can lead to incorrect practices and policies that are driven from the models we create and analyze. This sensitivity analysis to unmeasured confounding allows a researcher to assess the strength of the effects in the presence of confounding. If the sensitivity analysis suggests that the results are not robust to unmeasured confounding then it is possible to suggest that further research needs to be done with data collected on the confounder as well. If results do hold up under sensitivity analysis then there can be more confidence in the effects that have been estimated.

Furthermore we note that while we chosen to use simplified models with a single confounder in this paper these techniques can be applied to a model of any size and any path where the confounder lies. By breaking the model down into the smaller scenarios that we have discussed, we can use the sensitivity analysis techniques to asses the individual paths and correct for some of the bias that may be present.

# 9 Appendix

## 9.1 Arguments Concerning Table 1.

We use theory from causal diagrams here to demonstrate the points in the text. We refer the reader to Pearl's (2000) textbook for theory and terminology on causal diagrams. We will consider the scenarios in Table 1 and explain the biases that are present. With all of the models, $U$ has been left out of the actual analysis even though it is in the figure.

For the exposure-mediator confounder, shown in figure 3a, we see that the total and indirect effects are biased. We condition on $C$ in the analysis but this does not block all backdoor paths from $A$ to $Y$. The reason the total effect is biased is because there exists an unblocked backdoor path from $A$ to $Y$. That path would be $A \leftarrow U \rightarrow M \rightarrow Y$. If there was no effect of $A$ on $Y$ and of $A$ on $M$, then the total effect of $A$ on $Y$ would be 0, but with the path above left uncontrolled for our analysis this would show an effect of $A$ on $Y$. This is why the total effect would be biased. For the indirect effect in this scenario we see that even if there were not a path from $A$ to $M$, there would appear to be an indirect effect because of the path: $A \leftarrow U \rightarrow M \rightarrow Y$. However, the direct effect is unbiased in the presence of an exposure-mediator confounder (note that if there is no direct $A \leftarrow Y$ edge then $A$ will be independent of $Y$ conditional on $M$ even if the unmeasured variable $U$ is present).

For the exposure-outcome confounder, shown in figure 3b, we see that the total and direct effects are biased. For the total effect if there is no path from $A$ to $Y$ and from $A$ to $M$, then there would be no total effect from $A$ on $Y$. However because of $U$ there would be a path $A \leftarrow U \rightarrow Y$ giving rise to association. This would suggest there is an effect even when there is not one. For the direct effect we see that if there is no direct path from $A$ to $Y$, there is an open path $A \leftarrow U \rightarrow Y$ when conditioning on $M$. Again we would see an effect even when there is not one. For the indirect effect we can see that the only path from $A$ which goes through $M$ is $A \rightarrow M \rightarrow Y$ and the exposure-outcome confounder would not generate bias.

For the mediator-outcome confounder, shown in figure 3c, we see that the direct and indirect effects are biased. For the direct effect if we condition on $M$ we should block all paths from $A$ to $Y$ through $M$, in this case we condition on a collider and open up the path $A \rightarrow M \leftarrow U \rightarrow Y$ and would have association between $A$ and $Y$ conditional on $M$ even if there were no true effect. For the indirect effect we see that even if there were not an effect of $M$ on $Y$ we would see association between $M$ and $Y$ because of the path $M \leftarrow U \rightarrow Y$, and thus if we ignore the mediator-outcome confounder, we may find an indirect effect even when it is absent. For the total effect we see that all backdoor paths from $A$ to $Y$ are blocked by $C$ even if $U$ is present.

## 9.2 Theoretical Explanation for Bias of Additional Paths

In section 4 it was noted that in the presence of an unmeasured confounder, $U$, omitted from the LSEM that for any variable V that has an edge into the variable at the left most edge of the unmeasured confounder, the effect estimate of the edge from V to the variable on the right most edge of the unmeasured confounder will be biased. In the case of exposure-outcome confounding and mediator-outcome confounding all estimated edges into the outcome will be biased for all variables that have edges into the exposure or the mediator respectively. In the case of exposure-mediator confounding all estimated edges into the mediator will be biased for all variables with edges into the exposure as well.

We will consider exposure-mediator confounding first as in Figure 3a. When LSEM estimates the effect of $A \rightarrow M$ it conditions on $A$. This blocks the path $C \rightarrow A \rightarrow M$. However, because of $U$, $A$ is a collider and when we condition on it we open up the path $C \rightarrow A \leftarrow U \rightarrow M$. Thus even if there was no direct effect of $C$ on $M$ we would have a non-zero estimate for the $C \rightarrow M$ edge in a LSEM which ignored $U$. We next consider exposure-outcome confounding which is shown in figure 3b. When the LSEM estimates the effect of $A \rightarrow Y$ it conditions on $A$. This blocks the paths $C \rightarrow A \rightarrow Y$ and $C \rightarrow A \rightarrow M \rightarrow Y$. However because $A$ is again a collider it opens up the path $C \rightarrow A \leftarrow U \rightarrow Y$. This introduces a new unblocked path from $C$ to $Y$ and would bias our estimate of the effect for the edge, $C$ on $Y$.

Finally, we consider mediator-outcome confounding last which is shown in figure 3c. When the LSEM estimates the effect of $M \rightarrow Y$ it conditions on $M$. This blocks the path $C \rightarrow A \rightarrow M \rightarrow Y$. However, $M$ is now a collider on the path $C \rightarrow A \rightarrow M \leftarrow U \rightarrow Y$. This introduces a new path from $C$ to $Y$ and would bias our estimate of the effect for the edge of $C$ on $Y$.

## Acknowledgments

Research reported in this publication was supported by the National Institutes of Health under award numbers T32NS048005 and R01ES017876. The content is solely the responsibility of the authors and does not necessarily represent the official views of the National Institutes of Health.